# Negative Landau damping in bilayer graphene

Tiago A. Morgado[1], Mário G. Silveirinha[1,2*]

[1]Instituto de Telecomunicações and Department of Electrical Engineering, University of Coimbra, 3030-290 Coimbra, Portugal

[2]University of Lisbon, Instituto Superior Técnico, Avenida Rovisco Pais, 1, 1049-001 Lisboa, Portugal

E-mail: tiago.morgado@co.it.pt, mario.silveirinha@co.it.pt

## Abstract

We theoretically demonstrate that a system formed by two coupled graphene sheets enables a negative damping regime wherein graphene plasmons are pumped by a DC current. This effect is triggered by electrons drifting through one of the graphene sheets and leads to the spontaneous light emission (spasing) and wave instabilities in the mid-infrared range. It is shown that there is a deep link between the drift-induced instabilities and wave instabilities in moving media, as both result from the hybridization of oscillators with oppositely signed frequencies. With a thickness of few nanometers and wide spectral tunability, the proposed structure may find interesting applications in nanophotonic circuitry as an on-chip light source.

---

[*] To whom correspondence should be addressed: E-mail: mario.silveirinha@co.it.pt

Over the last decade, there has been a growing interest in nanoscale plasmonic light sources [1-15] that potentially can be incorporated on novel, highly-integrated, on-chip nanophotonic devices. Most of these proposals operate at near-infrared and visible frequencies. Recently, the interest in plasmonic light sources has been extended to the mid-infrared (mid-IR) and THz frequency spectrums. Mid-IR and THz sources may have important applications [16-18] in spectroscopy, chemical and biological sensing, and also in health- and security-related fields that include trace-gas detection and heat-signature sensing, among others.

A promising way to generate THz and mid-IR radiation is rooted in plasma wave instabilities. In particular, the Dyakonov-Shur (DS) [19-21] and Ryzhii-Satou-Shur (RSS) [22-23] instabilities may provide interesting opportunities in this context. These instabilities may occur in single-gate field-effect transistors (FETs) under the injection of a DC current. The DS instability arises from the Doppler-shift effect at asymmetric boundaries in the FET channel that leads to wave amplification at the drain [19-21], whereas the RSS instability occurs due to the transit-time effect of fast-moving electrons in the high-field region of the drain side [22-23]. The emergence of DS and RSS instabilities with enhanced growth rate in a dual-grating-gate graphene FET were recently reported in Ref. [24]. DS-type instabilities were also recently demonstrated in a periodically gated two-dimensional plasmonic crystal under direct current [25].

A different type of plasma instabilities can be induced by electrons streaming through a neutralizing background of slow ions, or even by counter-streaming electrons and ions (or holes). Such wave instabilities, designated as "two-stream instabilities", were studied in gaseous plasmas [26-27] and also in solid-state plasmas (e.g., semimetals and semiconductors) [28-33]. The two-stream instabilities can be seen as the inverse effect of the Landau damping – the process in which plasma waves lose energy to charged particles. In fact, under certain conditions, the transfer of energy may occur in the opposite direction, i.e., from the moving charged particles to the plasma waves, leading to the emergence of exponentially growing oscillations [33]. The two-stream instabilities were recently investigated in a nonequilibrium system in which a stream of electrons with a well-defined quasi-momentum is injected into a doped graphene sample [34]. Furthermore, unstable plasma waves can also be generated from the interaction between drifting electrons and the elastic vibrations of the lattice ions (phonons) in high-mobility semiconductors [35-38].

In this Letter, we propose a novel and promising platform to generate tunable mid-IR radiation that relies on surface plasmon amplification (spasing [1]) triggered by drifting electrons in an ultra-compact structure formed by two nearby graphene sheets (bilayer graphene). It is shown that the drift-induced instabilities are characterized by a giant growth rate and result from the hybridization of the surface plasmon polaritons (SPPs) supported by the individual graphene sheets. Moreover, we demonstrate that these instabilities are intrinsically related to the wave instabilities in moving media [39-43]. In what follows, the time variation is assumed of the form $e^{-i\omega t}$, where $\omega$ is the oscillation frequency.

Figure 1 schematically illustrates the structure under study. It consists of two graphene sheets separated by a silicon (Si) gap with thickness $d$ and dielectric constant $\varepsilon_{r,\mathrm{gap}} = 12$ [44]. The outer dielectric regions are assumed to be formed by SiO$_2$ or h-BN with $\varepsilon_{r,s} \approx 4$ [44, 45], but the properties of these regions are not critical. The top graphene sheet is traversed by a DC electric current induced by a static voltage applied across the graphene sheet. Due to the high mobility of graphene, the drift velocity $v_0$ can be on the order of the Fermi velocity $v_F \approx c/300$ [46-47], where $c$ is the speed of light in vacuum.

Without the electron drift, the graphene sheets are characterized by a surface conductivity $\sigma_g(\omega)$, which is expressed by the Kubo formula [48-49] and accounts for both the intraband and interband contributions. It is supposed that at room temperature, $T = 300$ K, the intraband and interband scattering rates are $\Gamma_{intra} = 1/(0.35 \text{ ps})$ and $\gamma_{inter} = 1/(0.0658 \text{ ps})$, respectively [49]. In the supplementary materials [50], it is shown using the self-consistent field approach [51-53] (also known as random phase approximation) that in presence of drifting electrons the graphene conductivity can be written in terms of the conductivity with no-drift ($\sigma_g$) as:

$$\sigma_g^{drift}(\omega) \approx \frac{\omega}{\omega - k_x v_0} \sigma_g(\omega - k_x v_0). \tag{1}$$

In the above, $k_x$ denotes the wave number along the x-direction (the xoy-plane spatial variation of the fields is $e^{ik_x x}$) and it is assumed that the in-plane electric field is directed along x (longitudinal excitation). Thus, an electron drift leads to a Doppler shift such that the surface conductivity is evaluated at the frequency $\tilde{\omega} = \omega - k_x v_0$ [33, 35]. In addition, the no-drift conductivity needs to be multiplied by the pre-factor $\omega/\tilde{\omega}$. This property has remarkable consequences. Indeed, in the absence of a drift the passivity of the material response requires that $\text{Re}\{\sigma_g\} > 0$ for all frequencies. Crucially, with a drift current one has $\text{Re}\{\sigma_g^{drift}\} = \text{Re}\{\sigma_g \omega/\tilde{\omega}\}$, and hence for sufficiently large $v_0$ the Doppler shifted frequency may flip its sign, $\tilde{\omega} < 0$, and lead to $\text{Re}\{\sigma_g^{drift}\} < 0$, i.e., to optical gain. In these circumstances, the graphene plasmons may gain energy at the cost of the kinetic energy of the streaming current carriers, an effect which we will refer to as "negative Landau damping".

To understand the impact of the drifting electrons on the optical response, next we study the natural modes of oscillation of the system depicted in Fig. 1. To begin with, we note that a negative damping can only occur for rather large values of the wave number $k_x > \omega / v_0$. Thus, it is justified to neglect time-retardation effects and adopt a quasi-static approximation [50]. Within this approximation, the dispersion equation for the natural modes of oscillation in the cavity formed by the two graphene sheets can be written as

$$1 - \tilde{R} R e^{-2 k_\parallel d} = 0, \qquad (2)$$

with $k_\parallel = \sqrt{k_x^2} = |k_x|$. Here, $\tilde{R}$ and $R$ are the reflection coefficients at the top and bottom interfaces of the Si gap region, respectively, for a transverse magnetic (TM) polarized plane wave with magnetic field along the *y*-direction ($\mathbf{H} = H_y(x,z)\hat{\mathbf{y}}$). The reflection coefficient is given by [50]:

$$R(k_x, \omega) = \frac{k_\parallel + (\varepsilon_{r,\text{gap}} - \varepsilon_{r,s}) \kappa_g}{k_\parallel - (\varepsilon_{r,\text{gap}} + \varepsilon_{r,s}) \kappa_g}, \qquad (3)$$

with $\kappa_g(\omega) = i\omega\varepsilon_0 / \sigma_g(\omega)$. Evidently, in case of an electron drift the same formula holds with $\kappa_g$ replaced by $\kappa_g^{\text{drift}}(\omega) = i\omega\varepsilon_0 / \sigma_g^{\text{drift}}(\omega)$. For simplicity, we neglect the frequency dispersion of the dielectric materials (Si, SiO$_2$). In these conditions, the dependence of $R$ on frequency is fully determined by the parameter $\kappa_g(\omega)$. In particular, from Eq. (1) it follows that the reflection coefficient with an electron drift ($\tilde{R}$) is simply related to the reflection coefficient with no drift ($R$) as $\tilde{R} = R(\tilde{\omega})$.

Let us first consider the scenario wherein the gap distance is very large ($d \to \infty$). In such a case, the natural modes of a single graphene sheet without (with) a drift current correspond to the poles of $R$ ($\tilde{R}$). Let $\omega \equiv \omega' + i\omega'' = \omega_n(k_x)$ denote the dispersion of some branch of natural modes without a drift current. Clearly, the intrinsic damping of graphene's response implies that for $k_x$ real-valued one has $\text{Im}\{\omega_n(k_x)\} < 0$, corresponding to an oscillation with finite lifetime ($e^{-i\omega t} = e^{-i\omega' t} e^{\omega'' t}$ decays with time). Using the fact that $\tilde{R} = R(\tilde{\omega})$, it follows that the dispersion of the natural modes with a drift is $\omega_n^{\text{drift}}(k_x) = \omega_n(k_x) + k_x v_0$. Remarkably, this property shows that the natural modes with a drift current also satisfy $\text{Im}\{\omega_n^{\text{drift}}(k_x)\} < 0$ for $k_x$ real valued, i.e., even though the drift current can potentially lead to optical gain, the system remains stable and the natural modes lifetime is *identical* to that in the case with no drift. A simple way to understand this result is to make a parallelism with moving media [39-43]. The important point is that in the non-relativistic limit the optical response of a graphene sheet (with no drifting electrons) in translational motion with velocity $v_0$ is characterized by the Doppler shifted reflection coefficient $\tilde{R} = R(\tilde{\omega})$ [42, 54], i.e., has exactly the same response as a graphene sheet with drifting electrons. It is physically evident that a single body moving at a constant velocity cannot lead to wave instabilities [39-43]. Hence, given the similarity between our system and a moving material, it is understandable that a single graphene sheet with an injected current cannot lead to spasing.

The emergence of wave instabilities requires the interaction between two graphene sheets with *different* DC currents (which is the analogue of the interaction of two material bodies in shear motion [39-43]), e.g., the interaction between a graphene sheet with drifting electrons and a graphene sheet with no drifting electrons. To illustrate this, we depict in Fig. 2 the dispersion of the natural modes of the bilayer graphene structure sketched in Fig. 1. The dispersion diagram was obtained by numerically solving Eq. (2). In the numerical simulation it is assumed that the chemical potential of the graphene sheets is $\mu_c = 0.1 \text{ eV}$ and that $v_0 = v_F$. Figure 2(a) shows the complex oscillation frequency $f = f' + i f''$ ($f = \omega/(2\pi)$) as a function of $k_x$ for two different gap distances: $d = 5$ nm (blue curves) and $d = 2.5$ nm (green curves). Remarkably, the system supports not only natural oscillations with finite lifetime ($f'' < 0$), but also wave instabilities that grow exponentially with time ($f'' > 0$). This unstable regime emerges even considering the realistic material loss of the graphene sheets, and is due to the negative Landau damping provided by the drifting electrons. As expected, the growth rate increases as the gap distance $d$ decreases. This behavior is even more evident from the results of Fig. 2(b), where $f''$ is depicted as a function of the gap distance $d$. It can be seen that the growth rate is as high as $\omega'' \approx 2 \times 10^{13}$ s$^{-1}$ for $d = 5$ nm, increasing up to values of $\omega'' \approx 7 \times 10^{13}$ s$^{-1}$ as the gap distance approaches zero. Notably, this range of values for the growth rate is one order of magnitude larger than those provided by DS and RSS instabilities in a graphene FET [24] or by phonon-plasmon instabilities in high-mobility semiconductors [35-38]. For a gap distance $d > 9$ nm the system becomes stable, because the coupling between the plasmons supported by each graphene sheet becomes inefficient for large $d$. In the supplemental material [50], it is shown that the drift-induced wave instabilities are robust to variations of the drift velocity $v_0$.

The "bubble"-type dispersion diagrams in the unstable regime are characteristic of a spontaneous parity-time ($\mathcal{P} \cdot \mathcal{T}$) symmetry breaking [43, 55, 56]. Indeed, in the limit of no material loss the considered system is $\mathcal{P} \cdot \mathcal{T}$-symmetric, and in that case the dispersion diagram has mirror symmetry with respect to the real-frequency axis [43]. Moreover, in the limit of no material loss (i.e. $\text{Re}\{\sigma_g\} \to 0^+$) the growth rate of the unstable mode is maximal (not shown). An increase of the absorption loss of the graphene sheets always leads to a decrease in the strength of the instabilities. Indeed, the unstable response is *not* due to the conversion of loss into gain, but rather due to the conversion of kinetic energy of the drifting electrons into plasmon oscillations, or, from a purely electromagnetic point of view, due to the interaction between negative and positive frequency oscillators as further detailed next.

To unveil the origin of the unstable modes, the dispersion diagrams of the SPPs sustained by each individual graphene sheet are shown in Fig. 3, together with the hybridized modes of the bilayer graphene system. The SPP dispersion of the individual graphene sheets is determined by $k_\parallel - (\varepsilon_{r,gap} + \varepsilon_{r,s})\kappa_g = 0$ (sheet with no drifting electrons) and by $k_\parallel - (\varepsilon_{r,gap} + \varepsilon_{r,s})\kappa_g^{drift} = 0$ (sheet with drifting electrons). As seen in Fig. 3, the dispersion diagrams of the SPPs supported by the individual graphene sheets (solid and dashed blue lines) and the hybridized modes in the combined system (thick solid orange lines) intersect at the point $k_x = 2\omega'/v_0$ (red circle), which corresponds approximately to the frequency for which the growth rate $\omega''$ of the hybridized modes is maximal. Hence, the peak instability arises due to the hybridization of the surface plasmons supported by the individual graphene sheets with identical values of $(\omega, k_x)$, i.e., $\omega(k_{x,1}) = \omega^{drift}(k_{x,2})$ and $k_{x,1} = k_{x,2}$, which corresponds to the interaction of plasmons with the same energy and momentum. Furthermore, the Doppler shifted frequency $\tilde{\omega} = \omega - k_x v_0$ associated with the sheet with drifting electrons can be written as $\tilde{\omega} = \omega(k_x) - k_x v_0$ with $k_x = 2\omega(k_x)/v_0$, or equivalently, $\tilde{\omega} = -\omega(k_x)$ (for the red circle in Fig. 3). Thus, the negative Landau damping is effectively the result of the interaction of a positive frequency oscillator ($\omega > 0$) and a negative frequency oscillator ($\tilde{\omega} < 0$) [39, 42, 56-59], and implies a transfer of kinetic energy of the drifting electrons to the plasma wave. The described hybridization "selection rules", $\omega_1 = \omega_2$, $k_{x,1} = k_{x,2}$, and $\omega_1 \tilde{\omega}_2 < 0$, are actually rather general, and are coincident with the conditions required to induce an unstable response in moving systems [41, 43]. This further highlights the profound link between the two problems.

To gain further insight into the dynamics of the instabilities, we calculated the electromagnetic field distribution of the hybridized unstable mode [Fig. 4]. As seen, in Fig. 4(a) the fields associated with this mode are predominantly localized in the region within and near to the Si gap. Despite the evident absence of spatial symmetry along the *z* direction – caused by the effect of the drifting electrons – one can still recognize in Fig. 4(a) the surface plasmon character of the hybridized mode with the fields decaying exponentially away from the graphene sheets. Figure 4(b) depicts a density plot of the $E_x$-field on the *xoz* plane at a generic time instant, with the Poynting vector lines superimposed, both with (*i*) and without (*ii*) electron drifts in the top graphene sheet. As expected, Fig. 4(b)(*ii*) reveals that without drifting electrons the energy density flux (Poynting vector) is parallel to the direction of propagation (*x*-direction) in all points of space. Notably, with drifting electrons [Fig. 4(b)(*i*)], the Poynting vector gains a significant *z*-component perpendicular to the interfaces, due to the exchange of energy between the plasmons supported by the two interacting graphene sheets. As seen, the energy flows away from the graphene sheet with drifting electrons due to the conversion of kinetic energy into plasmon oscillations made possible by the negative Landau damping.

An exciting property of graphene is the possibility to dynamically tune its conductivity by means of chemical doping or with a gate voltage. Thus, tuning $\mu_c$ may allow tailoring the spectrum of the wave instabilities. Figure 5 depicts the real and imaginary parts of the unstable mode with highest growth rate ($\max\{f''(k_x)\}$) as a function of $\mu_c$ for the gap distances $d = 5$ nm and $d = 2.5$ nm. The results clearly demonstrate that the dominant "spasing" frequency, as well as its growth rate, can be controlled by adjusting the chemical potential. In particular, by varying $\mu_c$ in between 0.05 and 0.3 eV one can achieve enormous growth rates ($\omega''$ varies from $10^{13}$ s$^{-1}$ to almost $10^{14}$ s$^{-1}$) over a frequency spectrum that extends from 10 THz to 60 THz or even more. Therefore, these results suggest that the proposed plasmonic nanostructure may have quite interesting applications as a tunable mid-IR light source.

In summary, we have shown that a system formed by two nearby graphene sheets may become unstable and thereby may start spontaneously emitting mid-IR radiation when an electron drift is induced in one of the graphene sheets. It was demonstrated that the hybridization of the plasmons supported by the graphene sheets gives rise to wave oscillations that grow exponentially with time in the linear regime due to a negative Landau damping. The wave instabilities are rooted in the same principles as those characteristic of moving media, both resulting from the interaction of positive and negative frequency oscillators. The nanometer-scale characteristic dimensions, combined with the wideband tunability, make the proposed structure very attractive for applications in plasmonic circuitry as an integrated light source.

This work was partially funded by Fundação para a Ciência e a Tecnologia PTDC/EEI-TEL/4543/2014 and by Instituto de Telecomunicações under project UID/EEA/50008/2013. T. A. Morgado acknowledges financial support by Fundação para a Ciência e a Tecnologia (FCT/POPH) and the cofinancing of Fundo Social Europeu under the Post-Doctoral fellowship SFRH/BPD/84467/2012.

# Figures

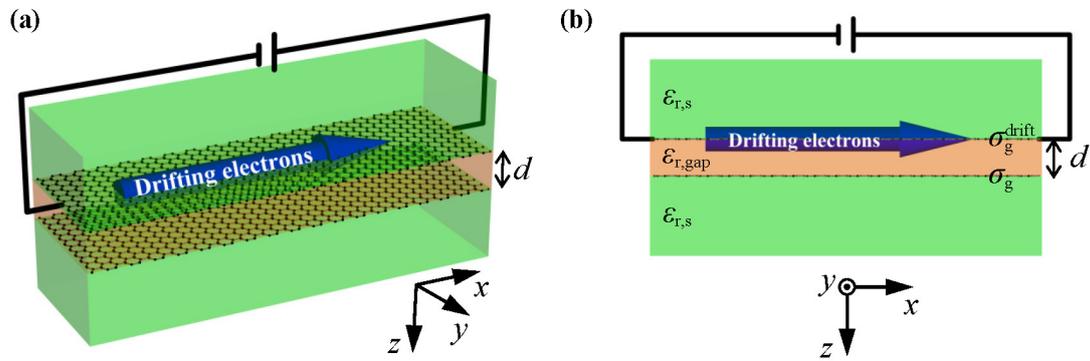

**Fig. 1.** Sketch of the system under study: two graphene sheets are separated by a silicon (Si) gap with thickness $d$. A voltage generator induces an electron drift in the top graphene sheet ($z = -0.5d$). (a) Perspective view; (b) front view.

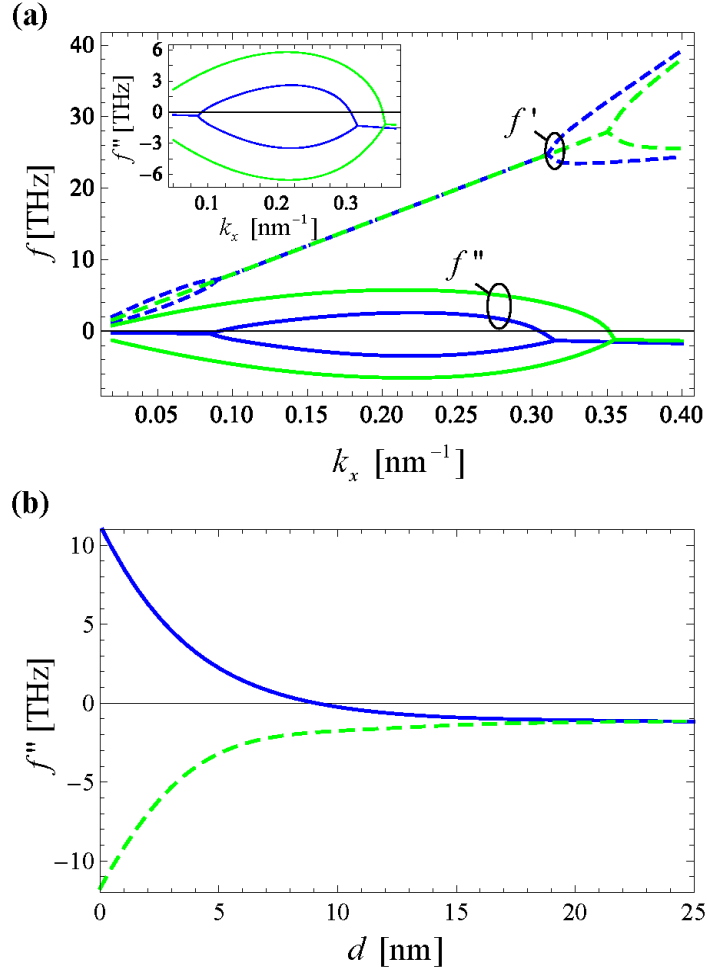

**Fig. 2.** (a) Real (dashed lines) and imaginary (solid lines) parts of the oscillation frequencies of the two relevant natural modes as a function of $k_x$ for (*i*) $d = 5$ nm (blue curves) and (*ii*) $d = 2.5$ nm (green curves). The inset shows a zoom-in of the imaginary parts of the oscillation frequencies. (b) Imaginary part of the oscillation frequencies as a function of the gap distance $d$. The transverse wave number is $k_x = 0.26$ nm$^{-1}$. The chemical potential is $\mu_c = 0.1$ eV.

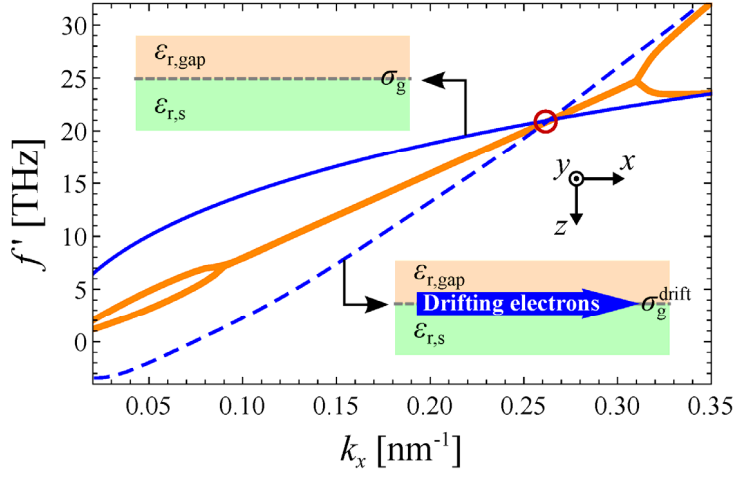

**Fig. 3.** Dispersion diagrams of the SPPs supported by the individual graphene sheets with $\mu_c = 0.1\,\text{eV}$. Solid blue line: without drifting electrons (bottom graphene sheet in Fig. 1); dashed blue line: with drifting electrons (top graphene sheet in Fig. 1). Orange thick line: dispersion diagram of the hybridized modes induced in the combined system with drifting electrons in the top graphene sheet [Fig. 1], for $d = 5\,\text{nm}$, and $\mu_c = 0.1\,\text{eV}$. The red circle indicates the point where $\omega' = 2\pi f' = k_x v_0 / 2$.

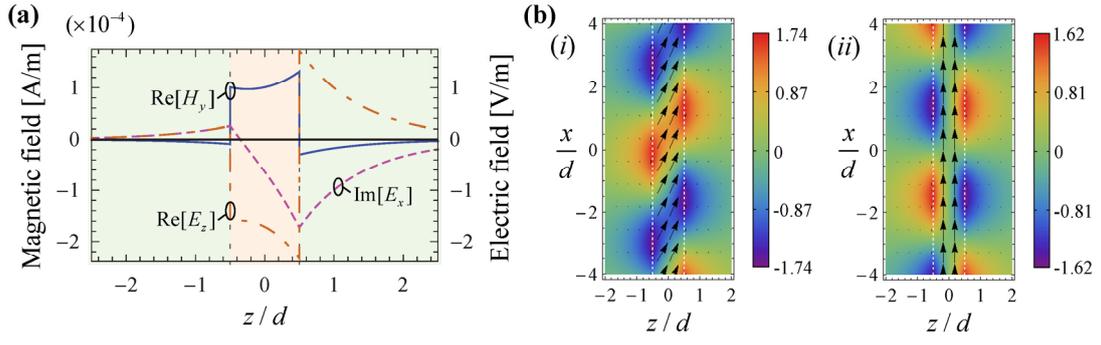

**Fig. 4.** (a) Electromagnetic field profiles of the unstable mode. The drifting electrons are in the graphene sheet located at $z = -0.5d$. The shaded region in the middle represents the Si gap and the vertical dashed lines indicate the locations of the graphene sheets. (b) Time snapshot of the *x*-component of the electric field $E_x$ (*i*) with, and (*ii*) without drifting electrons. The black arrows represent the Poynting vector lines. In both plots it is assumed that $k_x = 0.22$ nm$^{-1}$, $\mu_c = 0.1$ eV, and $d = 5$ nm.

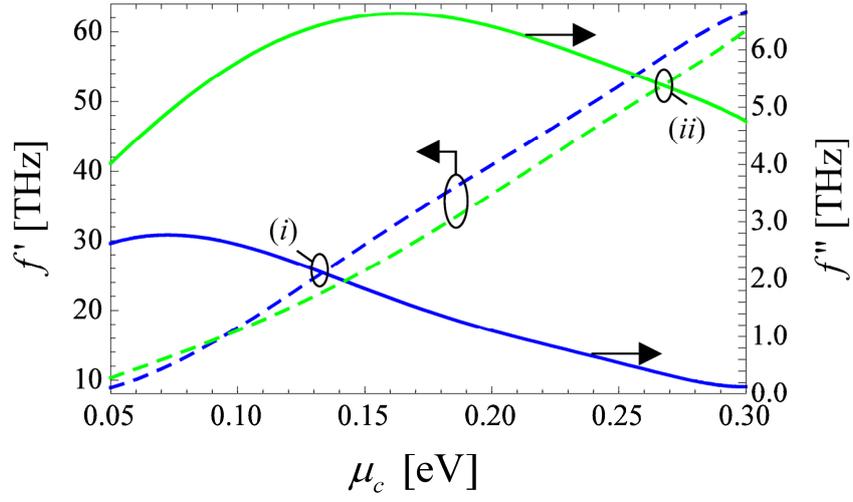

**Fig. 5.** Real (dashed lines) and imaginary (solid lines) parts of the oscillation frequency of the unstable mode with highest gain ($\max\{f''(k_x)\}$) as a function of the chemical potential $\mu_c$, assuming (*i*) $d = 5$ nm (blue curves) and (*ii*) $d = 2.5$ nm (green curves).

# Supplemental Material for the Manuscript

## "Negative Landau damping in bilayer graphene"

Tiago A. Morgado[1], Mário G. Silveirinha[1,2*]

[1]*Instituto de Telecomunicações and Department of Electrical Engineering, University of Coimbra, 3030-290 Coimbra, Portugal*

[2]*University of Lisbon, Instituto Superior Técnico, Avenida Rovisco Pais, 1, 1049-001 Lisboa, Portugal*

*E-mail:* tiago.morgado@co.it.pt, mario.silveirinha@co.it.pt

In this document we derive *a)* the longitudinal conductivity of graphene with drifting electrons, *b)* a closed-form expression for the reflection coefficient *R* of a plane wave incident on a graphene sheet, and *c)* we discuss the influence of the drift velocity $v_0$ on the wave instabilities.

## A. The longitudinal conductivity of graphene with drifting electrons

In what follows, we use the self-consistent field (SCF) approach, also known as random phase approximation [S1-S3], to determine the surface conductivity of graphene with drifting electrons. We adopt the following Hamiltonian to model the time dynamics of an electron in the presence of some perturbation (e.g., a static electric field) that originates a drift:

$$\hat{H} = \hat{H}_0 + \hat{\mathbf{p}} \cdot \mathbf{v}_0 + \hat{H}_{\text{int}}. \tag{S1}$$

In the above, $\hat{H}_0$ represents the two-dimensional massless Dirac Hamiltonian for the *K* point of graphene ($\hat{H}_0 = v_F \boldsymbol{\sigma} \cdot \hat{\mathbf{p}}$, with $\boldsymbol{\sigma}$ a tensor that depends on the Pauli matrices, $v_F$ is the Fermi velocity, and $\hat{\mathbf{p}} = -i\hbar\nabla$ the momentum operator), the term $\hat{\mathbf{p}} \cdot \mathbf{v}_0$ models a drift with a velocity $\mathbf{v}_0$, and $\hat{H}_{\text{int}} = V(\mathbf{r}, t)$ represents the time-dependent interaction due to the AC electric field. Note that the velocity operator is $\hat{\mathbf{v}} = \dfrac{\partial \hat{H}}{\partial \hat{\mathbf{p}}} = \dfrac{\partial \hat{H}_0}{\partial \hat{\mathbf{p}}} + \mathbf{v}_0$, i.e., it is the standard velocity operator in graphene (with no drift) plus a constant term ($\mathbf{v}_0$) determined by the drift velocity. The time-dependent potential is assumed of the form $V(\mathbf{r},t) = V_{\mathbf{q},\omega} e^{i\mathbf{q}\cdot\mathbf{r}} e^{-i\omega t} + c.c.$ where $\mathbf{q} = q_x \hat{\mathbf{x}} + q_y \hat{\mathbf{y}}$ is the in-plane wave vector, which determines also the direction of the average electric field (longitudinal excitation).

---

[*] To whom correspondence should be addressed: E-mail: mario.silveirinha@co.it.pt

Within the SCF approach, the longitudinal conductivity of graphene can be written as [S3]

$$\sigma_{L,\mathbf{q},\omega}^{\mathrm{SCF}} = \dfrac{i\omega e^2}{q^2} P, \tag{S2}$$

where $-e$ is the electron charge, $q^2 = \mathbf{q}\cdot\mathbf{q}$, and $P = \dfrac{\rho_{\mathbf{q},\omega}}{-eV_{\mathbf{q},\omega}}$ is a linear response function that determines the relation between the complex amplitude of the induced surface charge density $\rho_{\mathbf{q},\omega}$ and the complex amplitude of the average potential $V_{\mathbf{q},\omega}$. The operator associated with the charge density is $\hat{\rho}_{\mathbf{q}} = \dfrac{1}{A}(-e)e^{-i\mathbf{q}\cdot\mathbf{r}}$ with $A$ the area of the graphene sample. Hence, it is possible to write:

$$\dfrac{-1}{e}\rho_{\mathbf{q}}(t) = \mathrm{tr}\left\{\dfrac{1}{A}e^{-i\mathbf{q}\cdot\mathbf{r}}\hat{\rho}(t)\right\}, \tag{S3}$$

with $\hat{\rho}(t) = \sum_n f(E_n^0)|n(t)\rangle\langle n(t)|$ the time dependent density matrix of the system. Here, $f$ is the Fermi-Dirac distribution, $E_n^0$ represent the energies of the stationary states $|n^0\rangle$ of $\hat{H}_0$. The states $|n(t)\rangle$ are such that $|n(t=0)\rangle = |n^0\rangle$ and $\hat{H}|n(t)\rangle = i\hbar\partial_t |n(t)\rangle$.

To understand the effect of the drift, it is convenient to do a translation in space and define $|\tilde{n}(t)\rangle = e^{+\frac{i}{\hbar}\hat{\mathbf{p}}\cdot\mathbf{v}_0 t}|n(t)\rangle$. Then,

$$i\hbar\partial_t|\tilde{n}\rangle = e^{+\frac{i}{\hbar}\hat{\mathbf{p}}\cdot\mathbf{v}_0 t}\left(i\hbar\partial_t|n\rangle - \hat{\mathbf{p}}\cdot\mathbf{v}_0|n\rangle\right) = e^{+\frac{i}{\hbar}\hat{\mathbf{p}}\cdot\mathbf{v}_0 t}\left(\hat{H}_0 + \hat{H}_{int}\right)|n\rangle$$

$$= e^{+\frac{i}{\hbar}\hat{\mathbf{p}}\cdot\mathbf{v}_0 t}\left(\hat{H}_0 + \hat{H}_{int}\right)e^{-\frac{i}{\hbar}\hat{\mathbf{p}}\cdot\mathbf{v}_0 t}|\tilde{n}\rangle \quad (S4)$$

$$= \left(\hat{H}_0 + \hat{H}_{int}\Big|_{\omega - \mathbf{q}\cdot\mathbf{v}_0}\right)|\tilde{n}\rangle$$

where $\hat{H}_{int}\Big|_{\omega-\mathbf{q}\cdot\mathbf{v}_0} = V_{\mathbf{q},\omega}e^{i\mathbf{q}\cdot\mathbf{r}}e^{-i(\omega-\mathbf{q}\cdot\mathbf{v}_0)t} + c.c.$ represents an excitation with the Doppler shifted frequency $\tilde{\omega} = \omega - \mathbf{q}\cdot\mathbf{v}_0$. Using the property $e^{+\frac{i}{\hbar}\hat{\mathbf{p}}\cdot\mathbf{v}_0 t}e^{-i\mathbf{q}\cdot\mathbf{r}}e^{-\frac{i}{\hbar}\hat{\mathbf{p}}\cdot\mathbf{v}_0 t} = e^{-i\mathbf{q}\cdot\mathbf{r}}e^{-i\mathbf{q}\cdot\mathbf{v}_0 t}$, it is possible to rewrite Eq. (S3) as:

$$\frac{-1}{e}\rho_{\mathbf{q}}(t) = \frac{1}{A}\sum_n f(E_n^0)\langle n(t)|e^{-i\mathbf{q}\cdot\mathbf{r}}|n(t)\rangle$$

$$= \frac{1}{A}\sum_n f(E_n^0)\langle \tilde{n}(t)|e^{-i\mathbf{q}\cdot\mathbf{r}}|\tilde{n}(t)\rangle e^{-i\mathbf{q}\cdot\mathbf{v}_0 t} \quad (S5)$$

Crucially, from Eq. (S4) the dynamics of $|\tilde{n}\rangle$ is described by the Doppler shifted interaction Hamiltonian $\hat{H}_{int}\Big|_{\omega-\mathbf{q}\cdot\mathbf{v}_0}$. Hence, it is evident that the linear response ($P = \frac{\rho_{\mathbf{q},\omega}}{-eV_{\mathbf{q},\omega}}$) in presence of a drift is related to the linear response with no drift as:

$$P_{\mathbf{q},\omega}^{drift} = P_{\mathbf{q},\omega-\mathbf{q}\cdot\mathbf{v}_0} \quad (S6)$$

From Eq. (S2), this implies that:

$$\frac{\sigma_{L,\mathbf{q},\omega}^{SCF,drift}}{i\omega} = \frac{e^2}{q^2}P_{\mathbf{q},\omega}^{drift} = \frac{e^2}{q^2}P_{\mathbf{q},\omega-\mathbf{q}\cdot\mathbf{v}_0} = \frac{\sigma_{L,\mathbf{q},\omega-\mathbf{q}\cdot\mathbf{v}_0}^{SCF}}{i(\omega-\mathbf{q}\cdot\mathbf{v}_0)}. \quad (S7)$$

In the long wavelength limit $\sigma_{L,\mathbf{q},\omega}^{SCF}$ reduces to the standard local surface conductivity of graphene $\sigma_g(\omega) = \lim_{\mathbf{q}\to 0}\sigma_{L,\mathbf{q},\omega}^{SCF}$ (often referred to as the Kubo formula) [S3]. Hence, using the long wavelength approximation $\sigma_g(\omega) \approx \sigma_{L,\mathbf{q},\omega}^{SCF}$ we obtain the result of the main text:

$$\sigma_{\text{g}}^{\text{drift}}(\omega) = \frac{\omega}{(\omega - \mathbf{q} \cdot \mathbf{v}_0)} \sigma_{\text{g}}(\omega - \mathbf{q} \cdot \mathbf{v}_0). \tag{S8}$$

## *B. Derivation of the reflection coefficient*

The reflection coefficient for a transverse magnetic (TM) wave incident on a single graphene sheet (see Fig. S1) can be obtained in the usual way by expanding the electromagnetic field in all the regions in terms of plane waves, and then solving for the unknown wave amplitudes with mode matching. Assuming that the incident magnetic field is along the *y*-direction and has complex amplitude $H^{\text{inc}}$, it follows that the magnetic field in all space can be written as (the variation $e^{i(k_x x - \omega t)}$ of the fields is implicit):

$$\begin{aligned} H_y &= H^{\text{inc}} \left( e^{-\gamma_{\text{gap}} z} + R e^{\gamma_{\text{gap}} z} \right), \quad z < 0 \\ H_y &= H^{\text{inc}} T e^{-\gamma_s z}, \quad z > 0 \end{aligned} \tag{S9}$$

In the above, $\gamma_{\text{gap}} = \sqrt{k_x^2 - \varepsilon_{\text{r,gap}}(\omega/c)^2}$ and $\gamma_s = \sqrt{k_x^2 - \varepsilon_{\text{r,s}}(\omega/c)^2}$ are the propagation constants along *z* in the gap and substrate regions, *R* and *T* are the reflection and transmission coefficients, respectively.

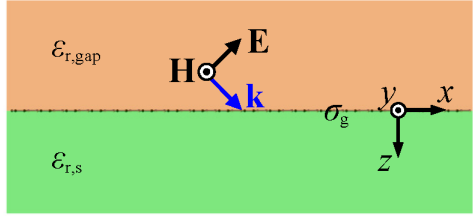

**Fig. S1.** Sketch of a single graphene sheet surrounded by different dielectrics.

The electric field associated with a generic plane wave $\mathbf{H} = H_0 e^{i\mathbf{k} \cdot \mathbf{r}} \hat{\mathbf{y}}$ is $\mathbf{E} = \frac{1}{-i\omega\varepsilon_0 \varepsilon_{\text{r},i}} \nabla \times \mathbf{H}$, and hence it is straightforward to decompose the electric field into plane waves, similar to Eq. (S9). By matching the tangential component of the electric field ($E_x|_{z=z_0^+} - E_x|_{z=z_0^-} = 0$) and by imposing the impedance boundary condition ($H_y|_{z=z_0^+} - H_y|_{z=z_0^-} = -\sigma_g E_x$) at the interface [S4], one can find that the reflection coefficient satisfies

$$R(k_x, \omega) = \frac{\kappa_g (\gamma_{gap} \varepsilon_{r,s} - \gamma_s \varepsilon_{r,gap}) - \gamma_s \gamma_{gap}}{\kappa_g (\gamma_{gap} \varepsilon_{r,s} + \gamma_s \varepsilon_{r,gap}) - \gamma_s \gamma_{gap}}, \quad (S10)$$

where $\sigma_g(\omega)$ is the graphene conductivity given by the Kubo formula [S4-S6] and $\kappa_g(\omega) = i\omega\varepsilon_0 / \sigma_g(\omega)$. In the quasi-static limit one has $\gamma_{gap} \approx \gamma_s \approx |k_x|$, and this gives the result of the main text.

## C. Effect of the drift velocity on the wave instabilities

Figure S2 depicts the real and imaginary parts of the oscillation frequency of the unstable mode (in a system formed by two coupled graphene sheets, as in Fig. 1 of the main text) as a function of $k_x$ and for drift velocities (a) $v_0 = 0.5 v_F$ and (b) $v_0 = 0.25 v_F$. The gap distance is $d = 1.25$ nm in (a) and $d = 0.33$ nm in (b). The chemical potential is taken equal to $\mu_c = 0.1$ eV in both cases.

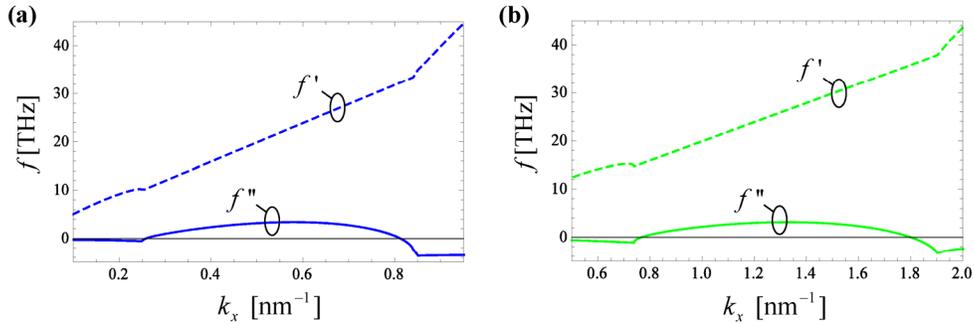

**Fig. S2.** (Color online) Real (dashed lines) and imaginary (solid lines) parts of the oscillation frequencies of the unstable mode as a function of $k_x$, assuming $\mu_c = 0.1$ eV. (a) $v_0 = c/600$ and $d = 1.25$ nm; (b) $v_0 = c/1200$ and $d = 0.33$ nm.

Figure S2 shows that the drift-induced instabilities are robust to variations of the value of the drift velocity, provided the gap distance can be made small enough. If $v_0$ is decreased the "selection rule" $\omega = -\omega + k_x v_0$ is only satisfied for larger values of $k_x$. Because a larger $k_x$ implies a smaller lateral decay length of the SPPs supported by each graphene sheet, the two individual SPPs only hybridize strongly for smaller gap distances *d*.